\documentclass[]{pasj00}
\usepackage{natbib}
\usepackage{times}

\usepackage{url}

\newcommand\jcap{{J. Cosmology Astropart. Phys.}}

\begin{document}
\title{Prospect for Future MeV Gamma-ray Active Galactic Nuclei Population Studies} 

\author{Yoshiyuki Inoue\altaffilmark{1}, Yasuyuki T. Tanaka\altaffilmark{2}, Hirokazu Odaka\altaffilmark{1,3}, Atsushi Takada\altaffilmark{4}, Yuto Ichinohe\altaffilmark{1,5}, Shinya Saito\altaffilmark{1,6}, Shin'ichiro Takeda\altaffilmark{1}, and Tadayuki Takahashi\altaffilmark{1}} 
\affil{$^1$Institute of Space and Astronautical Science JAXA, 3-1-1 Yoshinodai, Chuo-ku, Sagamihara, Kanagawa 252-5210, Japan}
\affil{$^2$Hiroshima Astrophysical Science Center, Hiroshima University, 1-3-1 Kagamiyama, Higashi-Hiroshima, Hiroshima 739-8526, Japan}
\affil{$^3$Max-Planck-Institut f\"{u}r Kernphysik, P.O. Box 103980, D 69029 Heidelberg, Germany}
\affil{$^4$Department of Physics, Graduate School of Science, Kyoto University, Kitashirakawa-Oiwakecho, Sakyo-ku, Kyoto 606-8502, Japan}
\affil{$^5$Department of Physics, Graduate School of Science, University of Tokyo, 7-3-1 Hongo, Bunkyo, Tokyo, 113-0033 Japan}
\affil{$^6$Department of Physics, Rikkyo University, 3-34-1 Nishi-Ikebukuro, Toshima-ku, Tokyo 171-8501, Japan}
\email{E-mail: yinoue@astro.isas.jaxa.jp}

\maketitle

\begin{abstract}
While the X-ray, GeV gamma-ray, and TeV gamma-ray skies have been extensively studied, the MeV  gamma-ray sky is not well investigated after the Imaging Compton Telescope (COMPTEL) scanned the sky about two decades ago. In this paper, we investigate prospects for active galactic nuclei population studies with future MeV gamma-ray missions using  recent spectral models and luminosity functions of Seyferts and flat spectrum radio quasars (FSRQs). Both of them are plausible candidates as the origins of the cosmic MeV gamma-ray background. If the cosmic MeV gamma-ray background radiation is dominated by non-thermal emission from Seyferts, the sensitivity of $10^{-12}\ {\rm erg\ cm^{-2}\ s^{-1}}$ is required to detect several hundred Seyferts in the entire sky. If FSRQs make up the cosmic MeV gamma-ray background, the sensitivity of $\sim4\times10^{-12}\ {\rm erg\ cm^{-2}\ s^{-1}}$ is required to detect several hundred FSRQs following the recent FSRQ X-ray luminosity function. However, based on the latest FSRQ gamma-ray luminosity function, with which FSRQs can explain up to $\sim30$\% of the MeV background, we can expect several hundred FSRQs even with the sensitivity of $10^{-11}\ {\rm erg\ cm^{-2}\ s^{-1}}$ which is almost the same as the sensitivity goal of the next generation MeV telescopes. 
\end{abstract}

\section{Introduction}
\label{intro}
The X-ray, GeV gamma-ray, and TeV gamma-ray skies have been deeply investigated by various telescopes. However, the MeV gamma-ray sky has not been extensively studied yet after the Imaging Compton Telescope (COMPTEL) onboard the {\it Compton Gamma-Ray Observatory} ({\it CGRO}) satellite scanned the sky in 1990s \citep{sch00_comptel}. This means that huge discovery space is buried in this energy range. 

Measurement of MeV gamma rays is hampered by huge background events. Under this difficulty, the COMPTEL achieved the sensitivity of $\sim10^{-10}\ {\rm erg\ cm^{-2}\ s^{-1}}$ in the MeV band and discovered $\sim$63 gamma-ray sources (32 of them were gamma-ray bursts) at 0.75--30 MeV \citep{sch00_comptel}. The COMPTEL has found 10 extragalactic active galactic nuclei (AGNs) and 5 unidentified high latitude objects \citep{sch00_comptel}. 

To open a new observational window in the MeV sky, various missions have been proposed. {\it ASTRO-H} \citep{tak14}, which will be launched in 2015, will have a sub-MeV instrument, the soft gamma-ray detector \citep[SGD,][]{taj10,wat12} covering at 40--600~keV. The SGD will provide precise spectra with higher sensitivity than previous instruments by introducing the concept of narrow-Field-of-View Compton telescope which can dramatically reduce background events \citep{tak01,tak04}. Other instruments are also proposed such as CAST \citep{nak12}, DUAL \citep{von12}, GRIPS \citep{gre12}, SMILE \citep{tak11}, COSI \citep{kie14}, and ASTROGAM\footnote{\url{http://astrogam.iaps.inaf.it/index.html}}. Moreover, the instrumental performance were tested with ballon experiments \citep{tak11,ban11}. These future missions will unveil the MeV sky in near future. In addition, a new image reconstruction algorithm that has improved sensitivity to multiple point-like sources has been proposed toward those future all-sky missions \citep{ike14}. The new era of the MeV gamma-ray astronomy is approaching.

A fundamental question to those missions is how many sources one can expect with their designed instruments. In the extragalactic sky, the dominant population in the MeV sky is still uncertain because the origin of the cosmic MeV gamma-ray background from 0.3~MeV to 100~MeV is not well established \citep[see][for a recent review]{ino14_cgb}. Seyferts which make up the cosmic X-ray background (CXB) cannot explain this MeV background because of the assumed cutoff at a few hundred keV \citep[e.g.][]{ued03,gil07}. Above 100~MeV, the total gamma-ray background is known to be predominantly composed of blazars \citep[e.g.][]{ino09,aje15}, star-forming galaxies \citep[e.g.][]{ack12_stb}, and radio galaxies \citep[e.g.][]{ino11}. Although the MeV background may comprises these GeV gamma-ray populations as well, the lower energy part of the MeV background spectrum is smoothly connected to the CXB spectrum and shows softer \citep[photon index of $\Gamma\sim$ 2.8][]{fuk75,wat97,wei00} than the GeV component \citep[photon index of $\Gamma\sim$ 2.4][]{ack15_cgb}, indicating their different origins. 

Several candidates have been considered as the origin of the MeV background. One was the nuclear-decay gamma rays from Type Ia supernovae \citep{cla75,zdz96,wat99}. However, the measurements of the cosmic Type Ia supernovae rates show that the rates are not enough to explain the observed flux \citep[e.g.][]{ahn05_sn,str05,hor10,rui15}. 
Seyferts may naturally explain the MeV background up to a few tens of MeV and the smooth connection to the CXB \citep[][hereinafter I08]{sch78,fie93,ste99,ino08}. Non-thermal electrons in coronae can generate the power-law spectrum in the MeV band after the thermal cut-off via the Compton scattering of disk photons (I08). 
Flat-spectrum radio quasars (FSRQs) whose peak in the spectrum locates at MeV energies \citep{blo95,sam06} are also expected to significantly dominate the MeV background \citep[][hereinafter A09]{aje09}. 
Radio galaxies have been also discussed as one of the origins of the MeV background \citep{str76}. Recent studies have revealed that both of lobe \citep{mas11} and core \citep{ino11} emissions from radio galaxies could contribute only $\sim10$\% of the MeV gamma-ray background flux.
Dark matters has also been discussed as the origin of the MeV background \citep[e.g.][]{ahn05_dm1,ahn05_dm2}. Those MeV mass dark matter particle candidates are less natural than GeV-TeV dark matter candidates. Therefore, Seyferts and FSRQs can be regarded as potential astrophysical origins of the MeV background.

The purpose of this paper is to investigate prospects of extragalactic observations by future MeV instruments, especially of statistical aspects of AGNs. We focus only on Seyferts and FSRQs in this paper. For this purpose, we use recent luminosity functions (LFs) and spectral models of Seyferts and FSRQs in literature. Since a quantitative estimate based only on the COMPTEL results is not easy due to the small number of detected sources, we adopt recent X-ray or gamma-ray luminosity functions and spectral models of Seyferts and FSRQs. 

This paper is organized as follows. MeV gamma-ray emission models of Seyferts and FSRQs are described In Section \ref{sec:sed}. LFs of Seyferts and FSRQs are presented in Section \ref{sec:lf}. In Section \ref{sec:count}, results of the expected source counts and redshift distributions are presented. Discussions and conclusions are given in Section \ref{sec:dis_con}. Throughout this paper, we adopt the standard cosmological parameters of $(h, \Omega_M , \Omega_\Lambda) = (0.7, 0.3, 0.7)$.

\section{Spectra of Active Galactic Nuclei in the MeV gamma-ray band}
\label{sec:sed}
\subsection{Seyferts}
The X-ray spectra of Seyfert are phenomenologically explained by a combination of following components: a primary power-law continuum with a cutoff at $\sim$0.3~MeV in the form of $E^{-\Gamma}\exp(-E/E_c)$, absorption by surrounding gas, emission lines, a reflection component, and a soft excess of emission at $\lesssim2$~keV \citep[e.g.][]{dad08}. Relative fractions of these components vary with sources. Physically Comptonization of disk photons in a corona above the accretion disk generate the primary power-law continuum \citep[see e.g.][]{kat76,poz77,sun80}.  The temperature of the corona roughly determines the position of the spectral cutoff and the photon index of the intrinsic continuum together with the optical depth \cite[see e.g.][]{zdz94}. The Compton reprocessed emission and bound-free absorption of the primary continuum by surrounding cold matter generate the reflection component \citep{lig88,mag95}.

\begin{figure}
\centering
\includegraphics[bb=50 30 410 282,width=8cm]{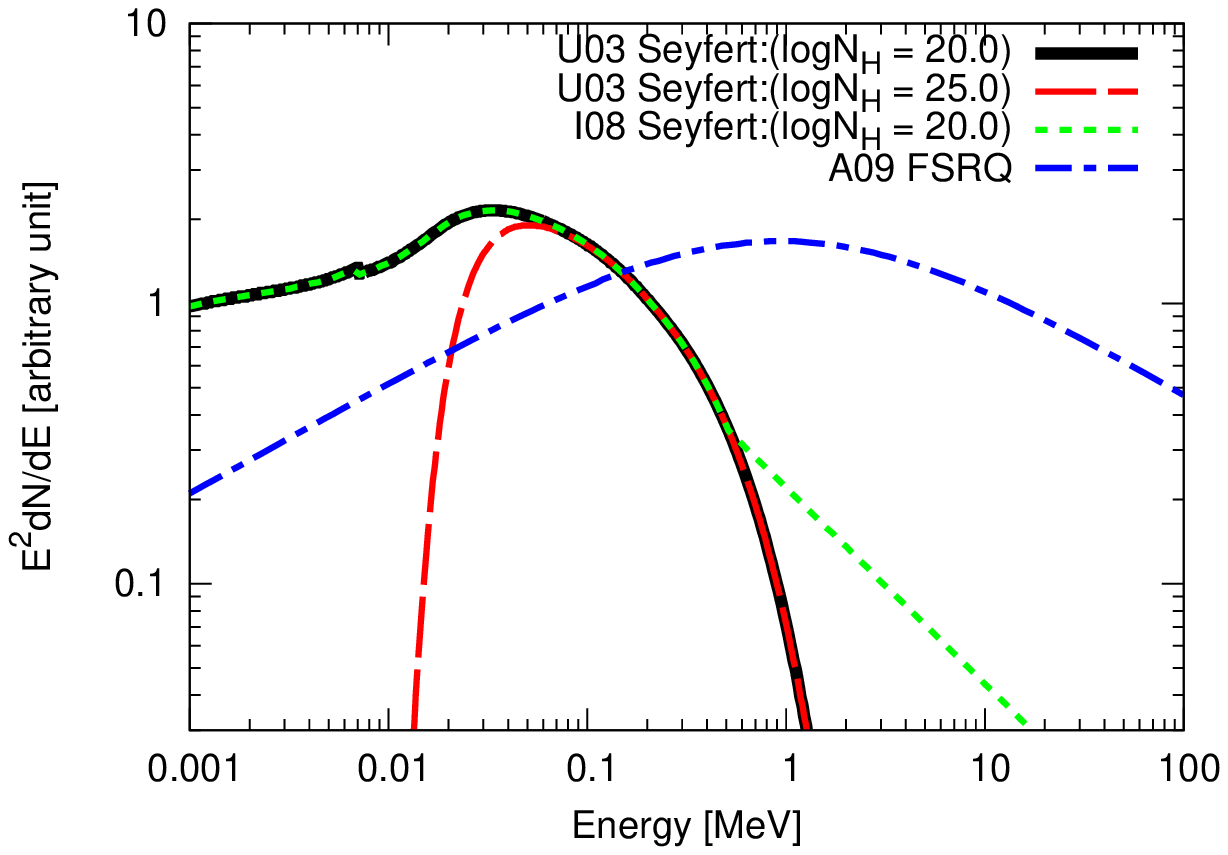} 
\caption{Spectral templates for Seyferts and FSRQs. Solid, dashed, dotted, and dot-dashed curve corresponds \citet[][U03]{ued03} Seyfert spectral model with a thermal cutoff at 0.3~MeV with $N_H=10^{20}\ {\rm cm^2}$, U03 Seyfert spectral model with a thermal cutoff at 0.3~MeV with $N_H=10^{25}\ {\rm cm^2}$,   \citet[][I08]{ino08} Seyfert spectral model with a non-thermal tail in the MeV band with $N_H=10^{20}\ {\rm cm^2}$, and \citet[][A09]{aje09} FSRQ spectral model having a peak at 1~MeV. The shape is assumed to be independent of luminosities for these models. \label{fig:sed_1} }
\end{figure}

Although coronae with temperature of $\sim0.1$~MeV do not produce significant MeV gamma-ray emission, some spectral models predict a power-law tail after the thermal cutoffs (e.g. I08). If the corona is composed of thermal and non-thermal populations, a MeV tail will appear after the cut-off. Such non-thermal electrons may exist if magnetic reconnection heats the corona \citep{liu02} as in the Solar flares \citep[e.g.][]{shi95} and Earth's magnetotail \citep{lin05}. The MeV background can be explained by the Seyferts explaining the CXB which have non-thermal electrons in the coronae having $\sim4$\% of the total electron energy  (I08). 

Observationally, the cutoff energy of Seyferts are constrained at $\gtrsim0.2$~MeV \citep{ric11}. The Oriented Scintillation Spectroscopy Experiment ({\it OSSE}) clearly detected emission up to 0.5~MeV in the spectrum of the brightest Seyfert NGC~4151 \citep{joh97}. This measurement constrained non-thermal fraction to be $\lesssim$15\% \citep{joh97}. Interestingly, future radio observations are able to probe those non-thermal tail through synchrotron emission \citep{lao08,ino14}. 

In this paper, as shown in Figure~\ref{fig:sed_1}, two primary spectral models are applied for Seyferts as in \citet{ino13}. One is thermal spectral model having a power-law continuum with a cutoff \citep[see e.g.][hereinafter U03]{ued03}. We adopt $\Gamma=1.9$ and $E_c=0.3$~MeV.  The other is thermal plus non-thermal spectral model (see I08 for details) which could explain the MeV background spectrum. We adopt the same parameters as in I08 with the non-thermal photon index of 2.8, but setting the cutoff at 0.3~MeV. For the Compton reflection component, we use a Compton reflection model \citep{mag95}, assuming a solid angle of $2\pi$, an inclination angle of $\cos i = 0.5$, and solar abundance for all elements. To calculate absorbed spectra, we use an absorption model called "wabs" developed for the XSPEC package. 

\subsection{FSRQs}

\begin{figure}
\centering
\includegraphics[bb=50 30 410 282,width=8cm]{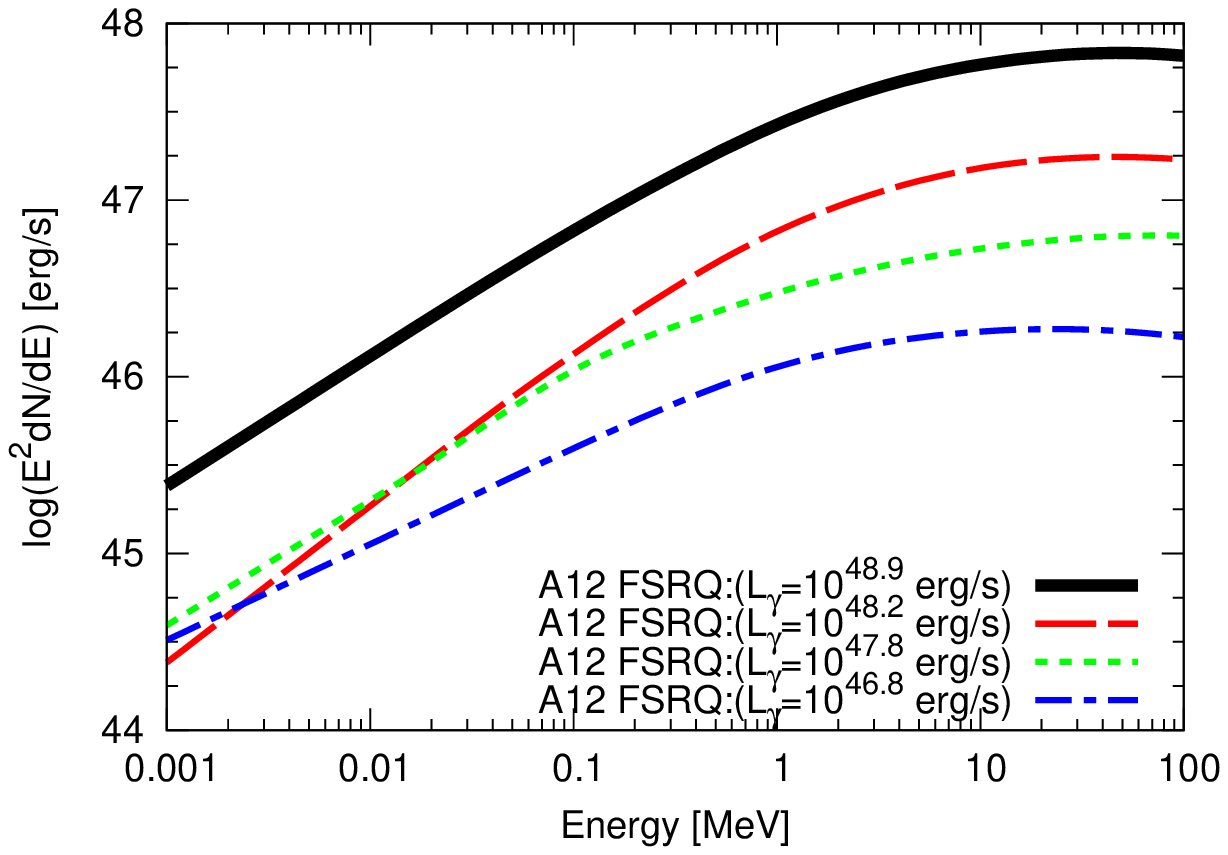} 
\caption{Spectral templates for FSRQs, but for A12 models. Solid, dashed, dotted, and dot-dashed curve corresponds to the 0.1-100~GeV luminosity of $L_{\gamma}=10^{48.9}$, $10^{48.2}$, $10^{47.8}$, and $10^{46.8}\ {\rm erg\ s^{-1}}$, respectively, based on \citet{aje12}. \label{fig:sed_2} }
\end{figure}

Multi-wavelength observations allowed us to study overall spectra of blazars from radio to gamma-ray. The observed blazar spectra consists of two broadband emission components \citep{ulr97,fos98,ghi98,kub98,abd10_blz}. One is synchrotron radiation component and the other is inverse Compton (IC) component in which electrons scatter internal synchrotron radiation \cite[see e.g.][]{jon74} or external radiation \cite[see e.g.][]{der93,sik94}. 

Blazars are divided into two categories by their optical spectra. Those are BL Lacertae objects (BL Lacs) and FSRQs. In the case of BL Lacs, the highest observable energy part of the synchrotron emission appears in the X-ray band. X-ray spectra of BL Lacs show soft spectra with a photon index of $\Gamma\sim2-3$.  In the case of FSRQs, the lowest observable energy part of the IC emission appears in the X-ray band showing harder spectra. Although BL Lacs are not expected to the dominant contributors of the MeV background, FSRQs may make up the whole MeV background (A09). Hence, in this paper, we do not consider BL Lacs but only FSRQs for blazars. 

We consider two spectral energy distribution (SED) models for FSRQs. One is based on the {\it Swift}-BAT FSRQs (A09), while the other is based on the  {\it Fermi}-LAT FSRQs \citet[][hereinafter A12]{aje12}. Following A09, the SED is given by a double power-law model:
\begin{equation}
\frac{dN}{dE}\propto\left[\left(\frac{E}{E_b}\right)^{\Gamma_1}+\left(\frac{E}{E_b}\right)^{\Gamma_2}\right]^{-1},
\end{equation}
where we set $E_b=1$ MeV, $\Gamma_1=1.6$, and $\Gamma_2=2.9$ following A09 (see Figure~\ref{fig:sed_1}). The average photon index of the {\it Swift}-BAT FSRQs is $1.6\pm0.3$ at 15--55 keV (A09). Although theoretical models predict a spectral turnover at the MeV band for FSRQs \citep[see e.g.][]{ino96}, $E_b$ and $\Gamma_2$ in the MeV band are observationally unconstrained. Hence, these parameters are set FSRQs to make up the MeV background. Here, the latest catalog of the {\it Fermi}-LAT sources reported detection of 467 FSRQs. The mean of their photon index above 0.1 GeV of those gamma-ray bright FSRQs is typically $\sim2.4$ \citep{ack15_3LAC}, which is harder than that of the MeV background. Combining with the {\it Fermi}-LAT and {\it Swift}-BAT detected FSRQs, the average SED is modelled in Fig. 9 of A12 which are shown in Figure~\ref{fig:sed_2}. We also adopt these SEDs as the other FSRQ template. However, as discussed in the next section, the whole MeV background is not be able to be explained by the A12 FSRQ model.

\section{Luminosity Functions of Active Galactic Nuclei}
\label{sec:lf}

\begin{table*}
\caption{The parameters of the AGN Luminosity Functions \label{XLF-parameters}}
\begin{center}
\begin{tabular}{cccc}
{} & \cite{ued03}  & \cite{aje09}  &  \cite{aje12}\\
{} & Seyfert  & FSRQ & FSRQ\\
{} & 2-10~keV & 15-55~keV & 0.1-100~GeV\\
\hline
$A$$^a$ & $50.4\pm3.3$  & $0.533\pm0.104$ & $0.0306\pm0.0023$\\
$\log_{10}$$L^*$$^b$ & $43.94_{-0.26}^{+0.21}$  & $44.0$ & $\log_{10}(0.84\pm0.49)$+48.0\\
$\gamma_1$ & $0.86\pm 0.15$  & - & $0.21\pm0.12$\\
$\gamma_2$ & $2.23\pm 0.13$  & $3.45\pm0.20$  & $1.58\pm0.27$ \\
$z_c^*$ & $1.9$  & - & $1.47\pm0.16$\\
$\log_{10}$$L_a$$^b$ & $44.6$  & - & -\\
$\alpha$ & $0.335\pm 0.07$  & - & $0.21\pm0.03$\\
$p_1$ & $4.23\pm0.39$  & $3.72\pm0.50$ & $7.35\pm1.74$\\
$p_2$ & $-1.5$  & $-0.32\pm0.08$ & $-6.51\pm1.97$\\
$\mu$ & - & - & $2.44\pm0.01$\\
$\sigma$ & - & - & $0.18\pm0.01$\\
\hline
\multicolumn{4}{l}{$^{\rm a}$: In units of $10^{-7} {\rm Mpc}^{-3}$.} \\
\multicolumn{4}{l}{$^{\rm b}$: In units of ${\rm erg/s}$.} \\
\end{tabular}
\end{center}
\end{table*}

To obtain the expected source counts and redshift distributions of AGNs in the MeV gamma-ray band, a LF is required. We briefly review LFs of Seyferts and FSRQs in this section.

\subsection{Seyferts}
The cosmological evolution of Seyferts is investigated by various X-ray surveys \citep[see e.g. U03;][]{has05,gil07,ued14}. Based on these surveys, the observed X-ray luminosity functions (XLFs) are known to be well represented by luminosity-dependent density evolution (LDDE) models \citep[see e.g.][and references therein]{ued14}. 

In this study, we follow the U03 LDDE XLF at 2-10~keV for the Seyfert models, since the I08 Seyfert model is based on U03. We note that U03 modelled the evolution of X-ray emitting AGNs. The comoving number density $\rho_X$ in the LDDE is
\begin{equation}
\rho_X(L_X, z,N_{\rm H}) = \rho_X(L_X,0)f(L_X,z)\eta(N_{\rm H};L_X, z),
\end{equation}
where $L_X$ is the X-ray luminosity at 2-10~keV before the absorption, $z$ is the redshift, and $N_{\rm H}$ is the neutral hydrogen column density. $\rho_X(L_X, 0)$ is the AGN XLF at present day characterized by the faint-end slope index $\gamma_1 $, the bright-end slope index $\gamma_2$, and the break luminosity $L^*$, as:
\begin{equation}
\rho_X(L_X,0)=A \left[ \left( \frac{L_X}{L^*} 
\right)^{\gamma_1} + \left( \frac{L_X}{L^*} \right)^{\gamma_2} \right]^{-1} \ ,
\end{equation}
where $A$ is the normalization parameter having a dimension of volume$^{-1}$. 

The density evolution function $f(L_X,z)$ is given as:
\begin{eqnarray}  
  f(L_X,z)=\left\{\begin{array}{ll}
      (1+z)^{p_1} & z \le z_c(L_X), \\
      (1+z_c(L_X))^{p_1}
      \left( \frac{1+z}{1+z_c(L_X)} \right)^{p_2} & z > z_c(L_X), \\
    \end{array}\right.
\end{eqnarray} 
where $z_c$ is the redshift of evolutionary peak, given as
\begin{eqnarray}
z_c(L_X)=\left\{\begin{array}{ll}
    z_c^* & L_X \ge L_a, \\
    z_c^*(L_X/L_a)^\alpha & L_X < L_a. \\
    \end{array}\right.
\end{eqnarray} 

The function $\eta(N_{\rm H};L_{\rm X},z)$ is the distribution of absorption column density given in the following form in the XLF \citep{ued03}:
\begin{eqnarray}  
\eta(N_{\rm H};L_{\rm X},z)
 &=\left\{\begin{array}{ll}
     2-\frac{5+2\epsilon}{1+\epsilon}\psi(L_{\rm X},z)&  20.0 \leq \log N_{\rm H} < 20.5, \\
	\frac{1}{1+\epsilon}\psi(L_{\rm X},z) & 20.5 \leq \log N_{\rm H} < 23.0, \\
	\frac{\epsilon}{1+\epsilon}\psi(L_{\rm X},z) & 23.0 \leq \log N_{\rm H} < 24.0, 
    \end{array}\right.
\end{eqnarray} 
where  $\epsilon=$1.7 and $\psi(L_{\rm X},z)={\rm min}\{\psi_{\rm max}, {\rm max}[0.47-0.1(\log L_{\rm X} - 44.0), 0])\},$
for which $\psi_{\rm max}=({1+\epsilon})/({3+\epsilon})$.
We note the integration of $\eta(N_{\rm H};L_{\rm X},z)$ between $20.0\le\log N_{\rm H}\le24.0$ is unity.

The parameters obtained by the fit to the observed data of X-ray AGNs in U03 are shown in Table \ref{XLF-parameters}. We set the minimum and maximum of the X-ray luminosity as $L_{X,\rm min}=10^{41.5}$ and $L_{X,\rm max}=10^{48.0}$ erg s$^{-1}$, respectively, the same as in U03. The redshift range is set to be $0.0 \le z \le 5.0$, otherwise noted. To explain the CXB, the fraction of the Compton thick AGNs between $24.0\le\log N_{\rm H}<25.0$ is set to be the same as that of the population at  $23.0\le\log N_{\rm H}<24.0$ following U03. 

Recently, \citet{air10} suggested another evolution model for AGNs, the luminosity and density evolution (LADE) model. Although the distribution of absorption column density is not available for the LADE, we can test that model assuming the same absorption column density distribution as in U03. The overall source counts in the MeV band significantly decrease by a factor of $\sim3$ at the flux range of $10^{-11}$--$10^{-12}\ {\rm erg\ cm^{-2}\ s^{-1}}$. However, the latest study of X-ray AGN evolution \citep{ued14} found that the AGN XLF is not well described with the LADE model but with LDDE model. When we adopt the latest LDDE model \citep{ued14}, the Seyfert source counts in the MeV band decrease by a factor of $\sim40$\% at the flux range of $10^{-11}$--$10^{-12}\ {\rm erg\ cm^{-2}\ s^{-1}}$. As the I08 Seyfert model is based on U03, we follow the U03 LDDE model in this paper.

\subsection{FSRQs}
In the GeV gamma-ray band, A12 and \citet{aje14} have recently shown that the gamma-ray luminosity functions (GLFs) of FSRQs and BL Lacs can be described by LDDE models based on the {\it Fermi}--LAT FSRQs and BL Lacs, which was indicated before the launch of {\it Fermi} \citep[e.g.][]{nar06,ino09}. With the A12 GLF model, FSRQs can explain up to $\sim30$\% of the MeV background. In the X-ray band, A09 have investigated the XLF of FSRQs using the 3-year {\it Swift}--BAT AGN survey data including 26 FSRQs. A pure luminosity evolution (PLE) model successfully reproduced the observed distribution of X-ray FSRQs. Assuming this PLE model and the spectral model in \S. 2.2, FSRQs explained the whole MeV background.

\begin{figure*}[t]
\centering
\includegraphics[bb=50 30 410 282,width=12cm]{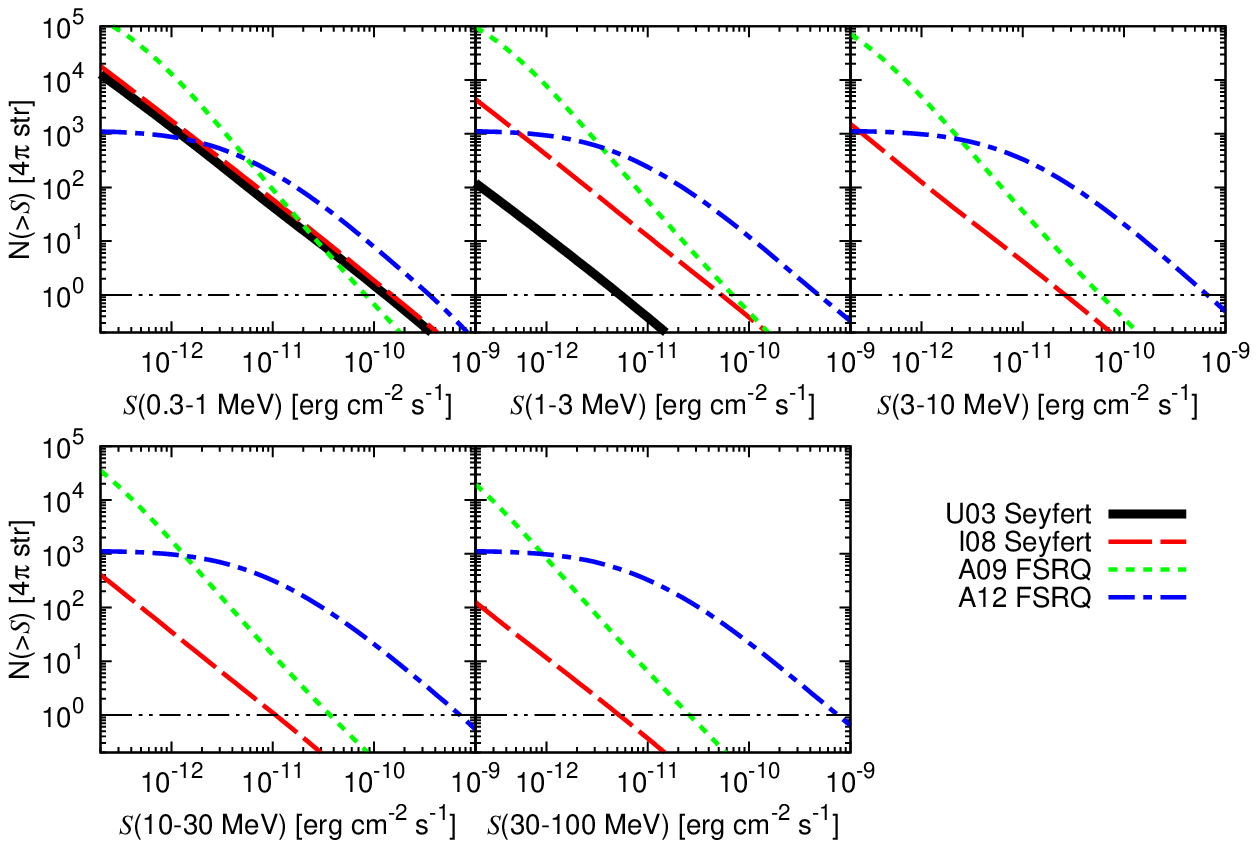} 
\caption{Cumulative source counts in the entire sky as a function of MeV gamma-ray flux (in $S$) of Seyferts and FSRQs. The five panels correspond to different photon energy bins as indicated in the panels. The solid, dashed, dotted, and dot-dashed curves are predictions based on \citet[][U03]{ued03} for Seyferts, \citet[][I08]{ino08} for Seyferts, \citet[][A09]{aje09} for FSRQs, and \citet[][A12]{aje12} for FSRQs. The horizontal thin double-dot-dashed line corresponds to one source detection level. Due to the thermal cutoff at 0.3~MeV in the spectral model of U03, detection of Seyferts above 3~MeV is not expected in the given flux ranges. 
\label{fig:count} }
\end{figure*}

These two recent FSRQ XLF and GLF models of A09 and A12 are adopted in this paper. The FSRQ XLF at 15--55~keV by A09 is given in the form of the PLE. The comoving number densities $\rho_X$ in the PLE is given as:
\begin{equation}
\rho_X(L_X, z) = \rho_X(L_X/e(z),0),
\end{equation}
where $L_X$ is the X-ray luminosity at 15-55~keV here. The local XLF is described as:
\begin{equation}
\rho_X(L_X,0)=A \left( \frac{L_X}{L^*} \right)^{-\gamma_2}.
\end{equation}
The evolution factor $e(z)$ is independent of luminosity given by:
\begin{equation}
e(z)=(1+z)^{p_1 +p_2z}
\end{equation}  
The values of the parameters for the A09 XLF are shown in Table \ref{XLF-parameters}. The minimum luminosity is also assumed to evolve with redshift as
\begin{equation}
L_{X, \mathrm{min}}(z)=L_{X, \mathrm{min,0}}\times e(z),
\end{equation}
where $L_{X, \mathrm{min,0}}$ is the minimum luminosity at $z=0$. We set $L_\mathrm{X,min,0}=3\times10^{44} \ {\rm erg \ s^{-1}}$, $L_{X,\rm max}$=10$^{50.0}$ erg s$^{-1}$, $z_{\rm min}=0.0$, and $z_{\rm max}=6.0$ (A09), otherwise noted. 

On the contrary, the A12 FSRQ GLF is given in the form of LDDE model. The comoving number densities $\rho_\gamma$ in the LDDE is given as:
\begin{equation}
\rho_\gamma(L_\gamma, z, \Gamma) = \rho_\gamma(L_\gamma,0)f(L_\gamma,z)\theta(\Gamma),
\end{equation}
where $L_\gamma$ is the gamma-ray luminosity at 0.1--100~GeV and $\Gamma$ is the intrinsic photon index between 0.1--100~GeV. $\rho_\gamma(L_\gamma, 0)$ is the FSRQ GLF at present. This is characterized as:
\begin{equation}
\rho_\gamma(L_\gamma,0)=A \left[ \left( \frac{L_\gamma}{L^*} 
\right)^{\gamma_1} + \left( \frac{L_\gamma}{L^*} \right)^{\gamma_2} \right]^{-1}.
\end{equation}
The density evolution function $f(L_\gamma,z)$ is given by:
\begin{equation}  
  f(L_\gamma,z)= \left[ \left( \frac{1+z}{1+z_c(L_\gamma)} \right)^{p_1} + \left( \frac{1+z}{1+z_c(L_\gamma)} \right)^{p_2} \right]^{-1}
\end{equation}
where $z_c$ is given as $z_c(L_\gamma)=z_c^* \cdot (L_\gamma/10^{48})^\alpha$

The function $\theta(\Gamma)$ describes the distribution of photon index, which is given by a Gaussian form in the GLF (A12):
\begin{equation}
\theta(\gamma)=\exp\left[-\frac{(\Gamma - \mu)^2}{2\sigma^2}\right], 
\end{equation} 
where $\mu$ and $\sigma$ are the mean and the dispersion of the Gaussian distribution. However, we adopt the average SED of FSRQs (see Fig. \ref{fig:sed_2}) not power-law SEDs. Therefore, we include the integrated $\theta(\Gamma)$ over $\Gamma$ between 1.8 and 3.0 which is independent of luminosity and redshift.

The parameters for the A12 GLF are shown in Table \ref{XLF-parameters}. The limits of gamma-ray luminosities, redshifts, and photon indices are set as $L_{\gamma,\rm min}$=10$^{44}$ erg s$^{-1}$,  $L_{\gamma,\rm min}$=10$^{52}$ erg s$^{-1}$, $z_{\rm min}$=0.01, and $z_{\rm max}$=6, the same as in A12, otherwise noted.

\section{Source Counts and Redshift Distribution}
\label{sec:count}
\begin{figure*}[t]
\centering
\includegraphics[bb=50 30 410 282,width=12cm]{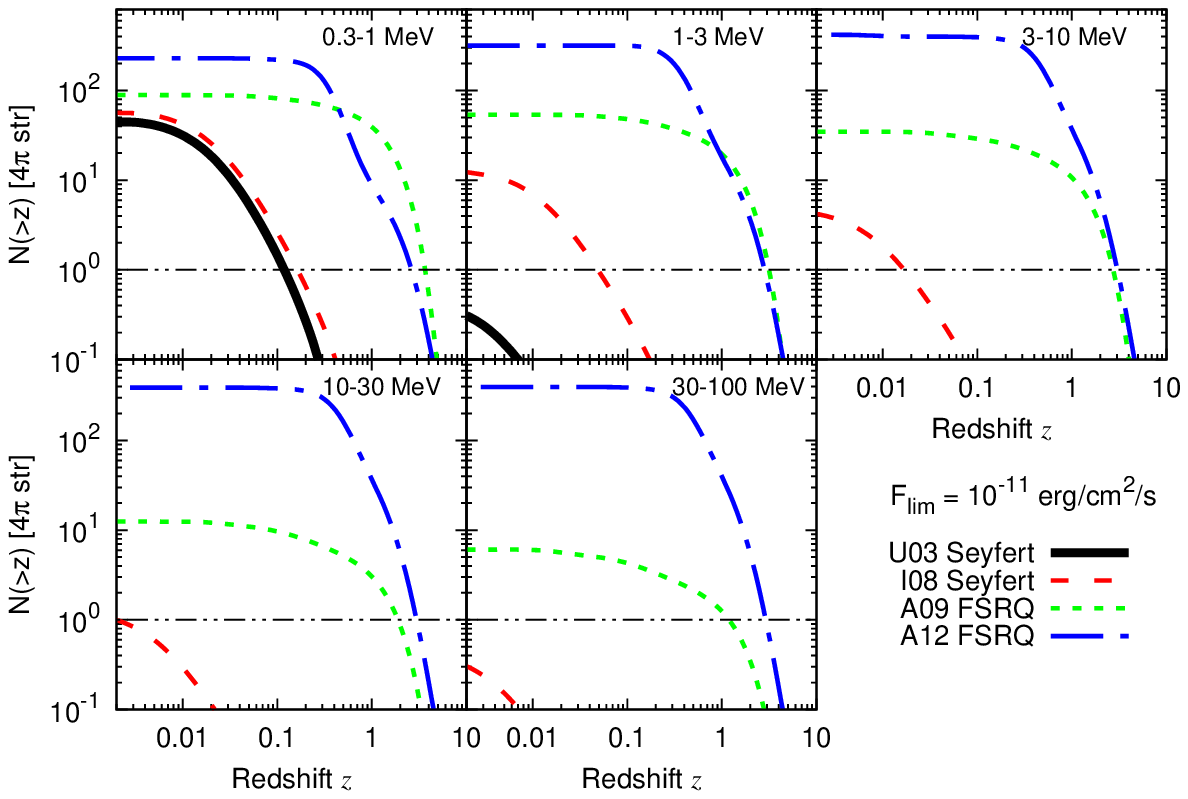} 
\caption{Expected cumulative redshift distribution of AGNs in the entire sky assuming the limiting sensitivity of $10^{-11}\ {\rm [erg\ cm^{-2}\ s^{-1}]}$. The five panels correspond to different photon energy bins as indicated in the panels. The solid, dashed, dotted, and dot-dashed curves are predictions based on \citet[][U03]{ued03} for Seyferts, \citet[][I08]{ino08} for Seyferts, \citet[][A09]{aje09} for FSRQs, and \citet[][A12]{aje12} for FSRQs. The horizontal thin double-dot-dashed line corresponds to one source detection level.\label{fig:z_1e-11} }
\end{figure*}

The cumulative source counts in the entire sky at a sensitivity of $S$, $N(>S)$, are given by integrating the luminosity function where the luminosity range is set to be $L_{\rm min}(S_{E_1, E_2},z)\le L\le L_{\rm max}$.  $S_{E_1,E_2}$ represents the sensitivity in a given energy range between $E_1$ and $E_2$ in the unit of ${\rm [erg\ cm^{-2}\ s^{-1}]}$ and $L_{\rm min}(S_{E_1, E_2},z)$ is the minimum detectable luminosity with a given sensitivity $S_{E_1,E_2}$ and a redshift $z$. Figure. \ref{fig:count} shows the cumulative source counts in the entire sky as a function of sensitivity limit at a given energy band in the entire sky predicted based on U03 for Seyferts with the thermal cutoff, I08 for Seyferts with the non-thermal tail, A09 for {\it Swift}--BAT FSRQs, and A12 for {\it Fermi}--LAT  FSRQs. The surface number density is given by dividing the number by $4\pi$~steradian. The energy bands are divided into five as 0.3--1~MeV, 1--3~MeV, 3--10~MeV, 10--30~MeV, and 30--100~MeV as indicated in panels. Hereinafter, we investigate required sensitivities for detection of several hundred sources. Several hundred samples may allow us to study cosmological evolution of a population \citep[see U03 for X-ray AGNs, A12 for gamma-ray FSRQs,][for gamma-ray BL Lacs]{aje14}. We also consider the sensitivity limit of $10^{-11}\ {\rm erg\ cm^{-2}\ s^{-1}}$ at the MeV band which is an order of magnitude better than the sensitivity of the COMPTEL and almost the same as the expected sensitivity of the next generation MeV instruments such as ASTROGAM.

Based on the U03 Seyfert model, $\sim20$~Seyferts at 0.3--1~MeV with the sensitivity limit of $10^{-11}\ {\rm erg\ cm^{-2}\ s^{-1}}$ are expected, which would be useful to investigate the position of the thermal cutoff energies of Seyferts. Once we can achieve the sensitivity limit of $10^{-12}\ {\rm erg\ cm^{-2}\ s^{-1}}$, $\sim1400$~Seyferts at 0.3--1~MeV. However, only $\sim10$~Seyferts are expected at 1--3~MeV. As the U03 Seyfert model relies on the thermal cutoff spectral model at 0.3~MeV, we can not expect any detection of U03-type Seyferts above several MeV.

If the non-thermal tails of Seyferts make up the MeV background (I08), more Seyferts at the MeV band would be expected. Based on the I08 Seyfert model, we can expect  $\sim60$, $\sim10$, $\sim5$, and a few Seyferts at 0.3--1, 1--3, 3--10, and 10--30~MeV, respectively, with the sensitivity limit of $10^{-11}\ {\rm erg\ cm^{-2}\ s^{-1}}$. By achieving the sensitivity limit of $10^{-12}\ {\rm erg\ cm^{-2}\ s^{-1}}$, we can expect a reasonable number of detections as $\sim1800$, $\sim400$, $\sim140$, $\sim40$, and $\sim10$ Seyferts at 0.3--1, 1--3, 3--10, 10--30, and 30--100~MeV, respectively. Therefore, the sensitivity limit of $\sim10^{-12}\ {\rm erg\ cm^{-2}\ s^{-1}}$ is required to detect several hundred Seyferts even if Seyferts dominate the MeV background radiation. 

If the origin of the MeV background is FSRQs (A09), we can expect $\sim90$, $\sim50$, $\sim30$, $\sim10$, and $\sim5$ FSRQs  at 0.3--1, 1--3, 3--10, 10--30, and 30--100~MeV, respectively, with the sensitivity limit of  $10^{-11}\ {\rm erg\ cm^{-2}\ s^{-1}}$. Once the sensitivity limit of $4\times10^{-12}\ {\rm erg\ cm^{-2}\ s^{-1}}$ is achieved, $\sim720$, $\sim420$, $\sim260$, $\sim90$, and $\sim40$ FSRQs  at 0.3--1, 1--3, 3--10, 10--30, and 30--100~MeV will be detected, respectively, because of strong positive evolution. However, based on the latest A12 FSRQ GLF, with which the whole MeV background can not be explained solely by FSRQs, the expected number will be $\sim220$, $\sim260$, $\sim390$, $\sim370$, and $\sim380$ FSRQs  at 0.3--1, 1--3, 3--10, 10--30, and 30--100~MeV, respectively, with the sensitivity limit of  $10^{-11}\ {\rm erg\ cm^{-2}\ s^{-1}}$. Moreover, at the sensitivity limit of the COMPTEL, $\sim10^{-10}\ {\rm erg\ cm^{-2}\ s^{-1}}$, about 10~FSRQs are expected based on the A12 GLF, which is consistent with the number of the COMPTEL blazars. The difference between FSRQ models mainly comes from the different evolutionary history.

When the sensitivity limit of $10^{-12}\ {\rm erg\ cm^{-2}\ s^{-1}}$ is achieved in the lowest energy band, expected detection rate for Seyferts and FSRQs will be one Seyfert per 22--28~deg$^2$ and one FSRQ per 4--40~deg$^2$, respectively, depending on assumed models. To avoid the source confusion, we may need angular resolutions of several degrees. ASTROGAM will achieve the FWHM of the angular resolution measure of $\sim4$~deg at 0.3~MeV \footnote{\url{http://astrogam.iaps.inaf.it/index.html}}. We note that COMPTEL had those of $\sim8$~deg at 0.5~MeV \citep{sch93}.

It is also important to know how distant AGNs can be observed by future missions. Since FSRQs have a spectral peak at the MeV band, future MeV missions will be able to detect distant FSRQs. In the GeV band, the most distant blazar confirmed by {\it Fermi} is PKS~0537-286 at $z=3.1$ at $\ge100$~MeV  \citep{ack15_3LAC}. \citet{rom04} and \citet{rom06} have reported one EGRET blazar candidate at $z\sim5.48$, but it has not been confirmed so far by {\it Fermi}. It is also expected that {\it Fermi} may eventually detect blazars at $z>6$ \citep{ino11_highz}. \citet{tak13} have reported a candidate {\it Fermi} gamma-ray blazar at $z\sim3-4$. 

\begin{figure*}[t]
\centering
\includegraphics[bb=50 30 410 282,width=12cm]{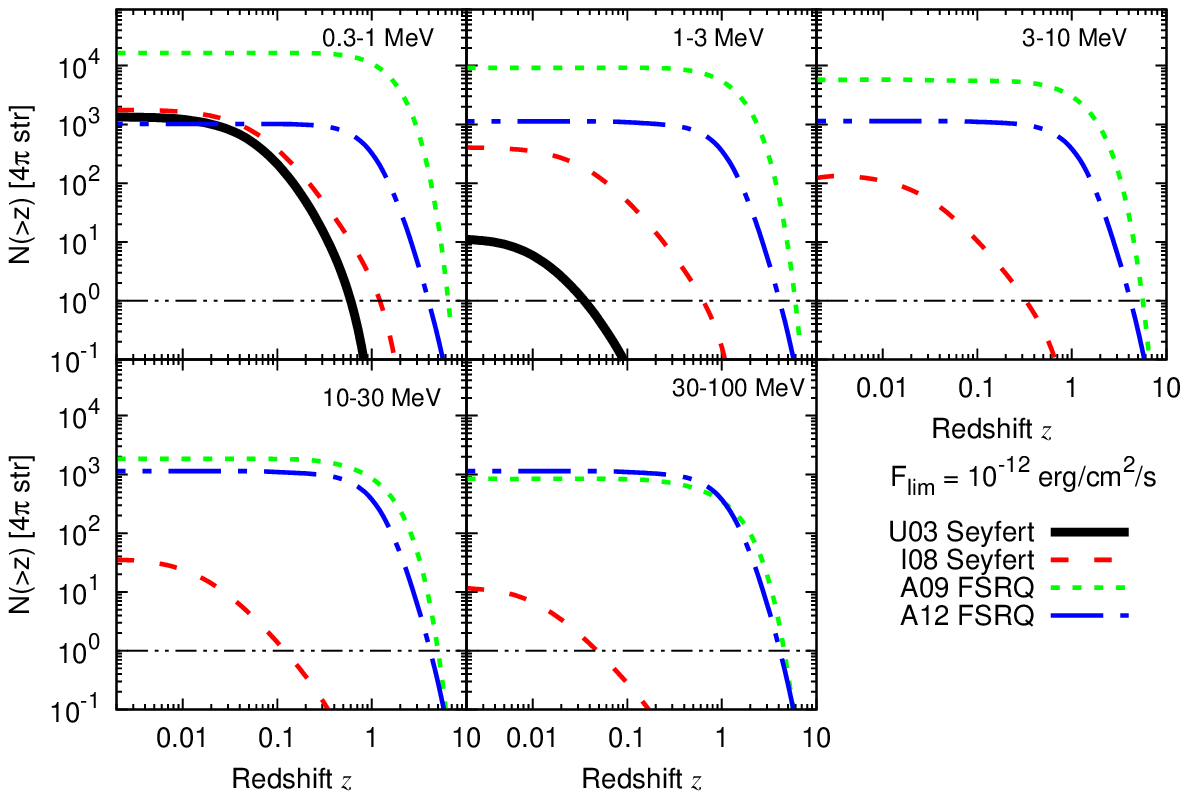} 
\caption{The same as Figure. \ref{fig:z_1e-11}, but for the limiting sensitivity of  $10^{-12}\ {\rm [erg\ cm^{-2}\ s^{-1}]}$. \label{fig:z_1e-12} }
\end{figure*}

Figure. \ref{fig:z_1e-11} shows the cumulative redshift distribution $N(>z)$ as a function of redshift at a given energy band in the entire sky assuming the sensitivity limit of $10^{-11}\ {\rm erg\ cm^{-2}\ s^{-1}}$ predicted based on U03 Seyfert model, I08 Seyfert model, A09 FSRQ model, and A12 FSRQ model. Here, we set $z_{\rm max}=8$ as the most distant known quasar is at $z=7.085$ \citep{mor11}. The energy bands were divided into five as 0.3--1~MeV, 1--3~MeV, 3--10~MeV, 10--30~MeV, and 30--100~MeV as indicated in panels. Figure. \ref{fig:z_1e-12} shows the same but assuming the sensitivity limit of $10^{-12}\ {\rm erg\ cm^{-2}\ s^{-1}}$.

For the Seyfert models, the detectable Seyferts will be only local Seyferts. Even if the sensitivity limit of $10^{-12}\ {\rm erg\ cm^{-2}\ s^{-1}}$ is achieved, the most distant Seyfert will be at $z\sim1$. On the other hand, for the FSRQ models, we can expect $z\sim3$--4 FSRQs with the sensitivity limit of $10^{-11}\ {\rm erg\ cm^{-2}\ s^{-1}}$. Once the sensitivity limit of $10^{-12}\ {\rm erg\ cm^{-2}\ s^{-1}}$ is achieved, $z\sim6$ FSRQs would be detectable even at the MeV band.

\section{Discussions and Conclusions}
\label{sec:dis_con}
In this paper, we studied the expected number counts and redshift distributions of AGNs (Seyferts and FSRQs) for future MeV missions based on recent AGN LFs and spectral models as Seyferts (I08) and FSRQs (A09) are discussed as the plausible origins of the cosmic MeV gamma-ray background. For Seyferts, we assume two primary spectral models. We considered a thermal spectral model with a cutoff at 0.3~MeV (U03) and a thermal plus non-thermal spectral model (I08). We adopt the (U03) LDDE XLF for both Seyfert models. For the thermal plus non-thermal spectral model (I08), a non-thermal component appears after the thermal cutoff at 0.3~MeV with the photon index of 2.8 to explain the MeV background up to a few tens of MeV. For FSRQs, we adopt two models. One is based on the {\it Swift}--BAT detected FSRQs (A09) and the other is based on the {\it Fermi}--LAT detected FSRQs (A12). The A09 FSRQ model can explain the entire MeV background solely by FSRQs, while the A12 FSRQ model makes up up to $\sim30$\% of the MeV background by FSRQs.

Since a thermal cutoff exists in spectra in the U03 Seyfert model, we can not expect any detections at $\gtrsim1$~MeV even with the sensitivity limit of $10^{-12}\ {\rm erg\ cm^{-2}\ s^{-1}}$. In contrast, if the origin of the MeV background is non-thermal tail from Seyferts (I08), we can expect several hundred Seyferts at the MeV band with the sensitivity limit of $10^{-12}\ {\rm erg\ cm^{-2}\ s^{-1}}$. Since each Seyfert is faint, we can detect only nearby ($z\lesssim1$) Seyferts even with the sensitivity limit of $10^{-12}\ {\rm erg\ cm^{-2}\ s^{-1}}$.

If FSRQs make up the whole MeV background (A09), the sensitivity of $\sim4\times10^{-12}\ {\rm erg\ cm^{-2}\ s^{-1}}$ is needed to detect several hundreds of FSRQs at the MeV gamma-ray band. However, based on the latest FSRQ GLF (A12), the sensitivity limit of  $10^{-11}\ {\rm erg\ cm^{-2}\ s^{-1}}$, which is almost the same as the expected sensitivity of the next generation MeV telescopes, would be enough to detect several hundreds of FSRQs. The difference between the two FSRQ models comes from the different evolutionary history.

Future MeV observational windows will range about three orders of magnitude in energy. With that wide energy range, the dominant population of the cosmic MeV gamma-ray background radiation would change with the energy. Furthermore, as there are uncertainties of luminosity functions and spectral models, the sensitivity of several times of $10^{-12}\ {\rm erg\ cm^{-2}\ s^{-1}}$ would be desirable to detect several hundred AGNs. 

It is important to know the origin of the MeV background because the expected source counts depends on it. To observationally unveil the origin of the MeV background, we need to resolve the sky into point sources as in soft X-ray. Here, the expected sensitivity of the next generation MeV instruments such as ASTROGAM is $\sim10^{-11}\ {\rm erg\ cm^{-2}\ s^{-1}}$. With this sensitivity, the expected resolved fraction will be a few percent of the total background in the I08 Seyfert scenario, while it will be $\sim$5\% in the A09 FSRQ scenario. Thus, the large fraction of the MeV sky will not be resolved even with the next generation instruments. However, as about one hundred to several hundred FSRQs are expected, we will be able to obtain typical MeV spectra and cosmological evolution of FSRQs and theoretically estimate their contribution to the MeV background more robustly. On the other hand, it would be hard for Seyferts since we can expect only about ten sources. 

The angular power spectrum of the MeV sky will provide a unique opportunity to unveil the origin of the MeV sky \citep{ino13}. The SGD onboard {\it ASTRO-H}  will be able to probe it up to 0.6~MeV. Although the angular power spectrum with the sensitivity of the SGD is dominated by the Poisson term $C_l^p$, a significant difference of the $C_l^p$ of Seyferts \citep{ino08} and FSRQs \citep{aje09} is  expected. This is because FSRQs are brighter but fewer than Seyferts. Anisotropies of the MeV sky would allow us to understand the origin of the MeV background and be useful to design future missions. 

\bigskip
The authors would like to thank Francesco Massaro for careful reading and constructive comments. The authors also thank Yasushi Fukazawa and Kazuhiro Nakazawa for useful comments. Y.I. acknowledges support by the JAXA international top young fellowship.

\end{document}